# DETECTION OF OPTICAL FLICKERING FROM THE SYMBIOTIC MIRA-TYPE BINARY STAR EF AQUILAE

V. V. DIMITROV[1,2], S. BOEVA[1], J. MARTÍ[3],
I. BUJALANCE-FERNÁNDEZ[3], E. SÁNCHEZ-AYASO[3], G. Y. LATEV[1],
Y. M. NIKOLOV[1], B. PETROV[1], K. MUKAI[4,5], K. A. STOYANOV[1],
R. K. ZAMANOV[1]

[1]*Institute of Astronomy and National Astronomical Observatory, Bulgarian Academy of Sciences, 72 Tsarigradsko Chaussee Blvd., 1784 Sofia, Bulgaria*
[2]*Department of Astronomy, Faculty of Physics, Sofia University "St. Kliment Ohridski", 5 J. Bourchier Blvd., 1164 Sofia, Bulgaria*
[3]*Departamento de Física (EPSJ), Universidad de Jaén, Campus Las Lagunillas, A3-420, E-23071, Jaén, Spain*
[4]*CRESST and X-ray Astrophysics Laboratory, NASA Goddard Space Flight Center, Greenbelt, MD 20771, USA*
[5]*Department of Physics, University of Maryland, Baltimore County, 1000 Hilltop Circle, Baltimore, MD 21250, USA*

E-mail: *vdimitrov@astro.bas.bg*

**Abstract.** We performed photometry with a 1 minute time resolution of the symbiotic stars EF Aquilae, AG Pegasi and SU Lyncis in Johnson *B* and *V* band. Our observations of the symbiotic Mira-type star EF Aql demonstrate the presence of stochastic light variations with an amplitude of about 0.25 magnitudes on a time scale of 5 minutes. The observations prove the white dwarf nature of the hot component in the binary system. It is the 11$^{th}$ symbiotic star (among more than 200 symbiotic stars known in our Galaxy) which displays optical flickering. For SU Lyn we do not detect flickering with an amplitude above 0.03 mag in *B* band. For AG Peg, the amplitude of variability in *B* and *V* band is smaller than 0.05 mag and 0.04 mag respectively.

## 1. INTRODUCTION

Symbiotic stars are long-period interacting binaries, consisting of an evolved giant transferring mass to a hot compact object. Their orbital periods are in the range from 100 days to more than 100 years. A cool giant or supergiant of spectral class G-K-M is the mass donor. The hot secondary component accretes material supplied from the red giant. In most symbiotic stars, the secondary is a degenerate star, typically a white dwarf, subdwarf (Mikołajewska 2003) or neutron star.

There are around 300 known symbiotic stars (Belczyński et al. 2000; Rodríguez-Flores et al. 2014). Among them flickering is detected in only 11 objects (including EF Aql), i.e. in 5% of the cases. On the basis of their infrared properties, the symbiotic stars are divided in three main groups: S-type, D-type, and D'-type (Allen 1982; Mikołajewska 2003). There are about 30 symbiotic stars classified as symbiotic Miras (Whitelock 2003). In three of them flickering is present, i.e. 10% of the objects. It seems that flickering can more often be detected in symbiotic Miras than among S and D'-type symbiotics.

## 2. OBSERVATIONS

The observations were performed with three telescopes equipped with CCD cameras:
- the 2.0 m RCC telescope of the Rozhen National Astronomical Observatory, Bulgaria (CCD cameras: VersArray 1300 B with 1340×1300 px resolution and 20 μm pixel, and Photometrics CE200A with 1024×1024 px resolution and 24 μm pixel);
- the 50/70 cm Schmidt telescope of Rozhen NAO (CCD camera FLI PL 16803, 4096×4096 px resolution, 9 μm pixel);
- the automated 41 cm Schmidt–Cassegrain telescope of the University of Jaén, Spain (CCD camera SBIG ST10-XME, 2184×1472 px resolution, 6.8 μm pixel; Martí, Luque-Escamilla & García-Hernández 2017).

## 3. FLICKERING OF EF AQL

EF Aql was identified as a variable star on photographic plates from Königstuhl Observatory (Reinmuth 1925).

Richwine et al. (2005) have examined the optical survey data for EF Aql and classify it as a Mira-type variable with a period of 329.4 d and amplitude of variability > 2.4 mag. Recently, Margon et al. (2016) reported that the optical spectrum shows prominent Balmer emission lines visible through at least H11 and [OIII] λ5007 emission. These emission lines and the bright UV flux detected in GALEX satellite images provide undoubted evidence for the presence of a hot companion as the M primary contributes negligible flux in this bandpass.

Thus EF Aql is classified as a symbiotic star, a member of the symbiotic Mira subgroup (Margon et al. 2016). Rapid aperiodic brightness variations, like the flickering from cataclysmic variables (Bruch 2000), is evident on all of our observations in *B* and *V* bands during 2016 and 2017. The flickering (stochastic photometric variations on timescales of a few minutes with amplitude of a few times 0.1 mag is a variability typical for the accreting white dwarfs in cataclysmic variables and recurrent novae. About the nature of the hot companion in EF Aql, Margon et al. (2016) supposed that the hot source is likely more luminous than a white dwarf, and thus may well be a subdwarf. The persistent presence of minute-timescale stochastic optical variations with the observed amplitude is a strong indicator that the hot component in EF Aql is a white dwarf.

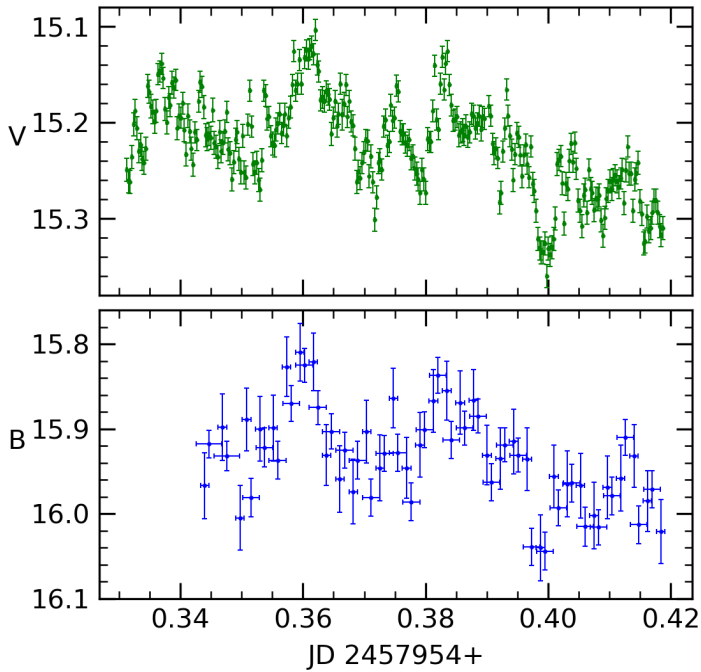

**Figure 1.** Simultaneous observations of EF Aql in *B* and *V* band. The flickering amplitude in *B* is 0.23 mag, and in *V* – 0.25 mag.

## 4. OBSERVATIONS OF AG PEG AND SU LYN

AG Peg is the slowest nova ever recorded. Its eruption began in 1855 and continued until 2001. Our observations from September 18, 2017 show that there is no flickering with amplitude above 0.05 in *B* band, and above 0.04 in *V* band.

SU Lyn was observed in X-rays by Mukai et al. (2016), who proposed that this system is a symbiotic star powered purely by accretion onto a white dwarf. The lack of shell burning leads to SU Lyn having very weak symbiotic signatures in the optical. On the night of January 24, 2018 we observed SU Lyn and we did not find any optical variability above 0.03 mag in *B* band.

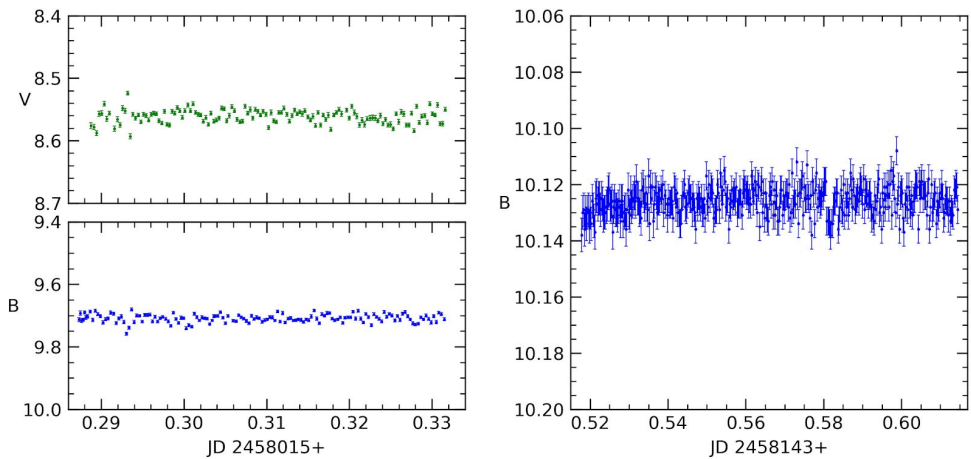

**Figure 2.** *Left:* simultaneous *B* and *V* band observations of AG Peg. No fast variability is detected. *Right:* observations of SU Lyn in *B* band. No fast variability is detected.

## 5. DISCUSSION

Most of the flickering-emitting symbiotics are bright X-ray sources. The X-ray emission is divided into five main types: α-type (massive WD with quasi-steady shell burning; Orio et al. 2007); β-type (collision of the winds of the WD and the RG; Muerset, Wolff & Jordan 1997); γ-type (X-rays from a neutron star accretor; Masetti et al. 2007); δ-type (X-rays, emitted from the boundary layer of the accretion disc; Luna et al. 2013); β/δ-type (two X-ray thermal components – soft from the colliding-wind region, and hard from the boundary layer). EF Aql has not been detected in X-rays yet, although it has still not been the subject of a pointed observation with imaging X-ray telescopes.

From the known flickering-emitting symbiotics, there are no α-types, because the energetics of the quasi-steady shell burning are higher than the flickering variability and the flickering becomes undetectable. γ-types are also not detectable, because neutron stars have not been observed as flickering sources.

Among the flickering-emitting symbiotic stars, Z And and o Cet are β-type sources, CH Cyg and MWC 560 are β/δ-type, RT Cru, T CrB and V648 Car are δ-type (Luna et al. 2013). Although RS Oph is considered a δ-type, it shows some peculiar features.

According to Sokoloski, Bildsten & Ho (2001), flickering is much less prevalent in symbiotic stars. The possible explanations are:
- the optical flux is dominated by light from the RG;
- the nebula "washes out" the flickering light;
- lack of an accretion disc;
- shell burning onto the WD surface.

Most probably SU Lyn does not show flickering activity because the flux is dominated by the red giant. In the case of AG Peg, the nova explosion has disrupted the accretion disc and currently the accretion must be spherically-symmetrical.

**Conclusions:** We observed optical flickering from the symbiotic star EF Aql with an amplitude of 0.23 mag in *B* band, and 0.25 mag in *V* band. The presence of flickering strongly suggests that the hot component is a white dwarf. Our observations of AG Peg and SU Lyn do not show variability above 0.05 mag.

**Acknowledgements:** This work was partially supported by the Program for career development of young scientists, Bulgarian Academy of Sciences (DFNP 15-5/24.07.2017), by grants DN 08-1/2016, DN 18-10/2017, DN 18-13/2017 (Bulgarian National Science Fund) and AYA2016-76012-C3-3-P from the Spanish Ministerio de Economía y Competitividad (MINECO).